# Intrinsic Fracture Nonreciprocity at the Nanoscale

Siwei Zhao,[1] Penghua Ying,[1,2] Guoqiang Zhang,[1] Ke Zhou,[3] Shengying Yue,[1] Yan Chen,[1, *] Yilun Liu[1, *]

[1] *Laboratory for multiscale mechanics and medical science, SV LAB, School of Aerospace, Xi'an Jiaotong University, Xi'an 710049, China*

[2] *Department of Physical Chemistry, School of Chemistry, Tel Aviv University, Tel Aviv 6997801, Israel*

[3] *College of Energy, SIEMIS, Soochow University, Suzhou 215006, China*

These authors contributed equally: Siwei Zhao, Penghua Ying.

* Corresponding authors.

E-mail: yanchen@xjtu.edu.cn (Y. Chen), yilunliu@mail.xjtu.edu.cn (Y. Liu)

**Abstract**

We reveal intrinsic fracture nonreciprocity, manifesting as directional asymmetry in crack resistance, in two-dimensional heterostructures engineered through lattice-mismatched interfaces. Density-functional theory combined with machine-learning molecular dynamics show that intrinsic lattice mismatch between bonded component crystals imprints asymmetric prestrain states at crack tips, governing bond-breaking thresholds through charge redistribution. The failure criterion obeys a universal exponential scaling law between normalized charge density and bond strain, insensitive to bonding chemistry and local atomic environment. The magnitude of nonreciprocity scales systematically with lattice mismatch, reaching 49% at 10% mismatch. Validation across hexagonal, square, rectangular, and oblique two-dimensional lattices confirms universality, establishing interface strain engineering as a general design principle that bridges electronic structure to nanoscale failure, enabling rational design of damage-tolerant nanostructures.

*Introduction.* — Fracture behaviors in crystalline materials have long assumed reciprocal, with crack propagation resistance identical regardless of loading direction [1]. This fundamental assumption underpins classical fracture mechanics and guides material design strategies [2]. However, the emergence of engineered interfaces at the nanoscale creates opportunities to break this reciprocity and realize direction-dependent fracture resistance—fracture nonreciprocity.

In analogy to a photodiode that permits current flow preferentially in one direction[3], fracture nonreciprocity could enable directional toughness control, but traditional toughening strategies through defect engineering [4], crack deflection [5], or bridging mechanisms [6], invariably compromise material integrity with irreversible structural modifications [7]. Recent advances in mechanical metamaterials have demonstrated macroscopic fracture nonreciprocity through geometric design[8-11], yet realizing directional fracture control at the atomic scale via intrinsic material properties, without artificial patterning, remains elusive.

In this letter, we demonstrate that coherent lattice-mismatched interfaces provide an intrinsic pathway to nanoscale fracture nonreciprocity, supported by density functional theory (DFT) calculations and molecular dynamics (MD) simulations driven by a dedicated machine learned potential (MLP). Using two-dimensional (2D) heterostructures as a model system, we reveal that lattice mismatch imprints asymmetric prestrain states at crack tips, establishing direction-dependent bond breaking thresholds. The effect is governed by a universal scaling relationship relating charge density to bond strain, independent of specific atomic composition or bonding characteristics. Validation across diverse symmetry 2D lattices confirms its universality and bridges electronic structure physics with mechanical failure, enabling rational design of damage-tolerant nanomaterials.

*Computational model and DFT calculations.* — We harness lattice-mismatched transition metal dichalcogenides (TMD) [12] heterostructures to achieve fracture nonreciprocity through interface strain engineering, preserving structural integrity. While TMDs heterostructures are well established for their tunable electronic and optical properties [3], their mechanical behavior remains largely unexplored. Their coherent interfaces therefore provide an ideal model to investigate the interplay between electronic structure and mechanical failure at the nanoscale.

We systematically investigate six representative TMD compositions derived from combinations of group-VI A transition metals ($M = \{Mo, W\}$) and group-VI B chalcogens ($X = \{S, Se, Te\}$) [Supplementary Materials (SM) Fig. S6], yielding 15 distinct lateral heterostructures [Fig. 1(a)]. Currently, ideal interfaces of lateral heterostructures can be successfully generated in experiments[13]. A fundamental structural feature emerges from the systematic variation in atomic radii: lattice constants $a$ increase monotonically with the atomic number of X while remaining nearly invariant with respect to M [SM Table S1]. Consequently, coalescing two TMDs with different $a$ introduces a lattice mismatch quantified as $\delta = (a_{max}-a_{min})/a_{max}$, where $a_{max}$ and $a_{min}$ are the larger and smaller lattice constants of the two constituent TMDs, respectively. Our systematic analysis identifies four distinct mismatch levels, $\delta \approx 0\%$, 4%, 7%, and 10% [SM Table S2]. This lattice mismatch significantly influences the atomic reconstruction at crack tips in TMDs heterostructures.

Quasi-static uniaxial-tension simulations were performed on lateral heterostructures with dimensions of approximately 29 Å (zigzag)× 58 Å (armchair), with interfaces centered along the armchair direction. Two pre-cracks, each 13 Å in length, were introduced symmetrically on opposite sides of each heterostructure [SM Fig. S8]. We simulate fracture by applying uniaxial tension incrementally along the armchair direction by uniformly scaling the cell length in 1% strain increments (relative to the pristine configuration), followed by full atomic relaxation, using DFT as implemented in the Vienna ab initio Simulation Package (VASP)[14-16]. Exchange correlation was described by the Perdew–Burke–Ernzerhof (PBE) functional within the generalized gradient approximation (GGA)[17-19] (see SM section S2 for details). Owing to residual misfit strain, crack-tip morphologies differed between regions: the larger-$a$ side exhibits contraction while the smaller-$a$ side undergoes opening. Comparative calculations were also performed for pristine $MX_2$ monolayers with identical crack configurations and for defect-free counterparts.

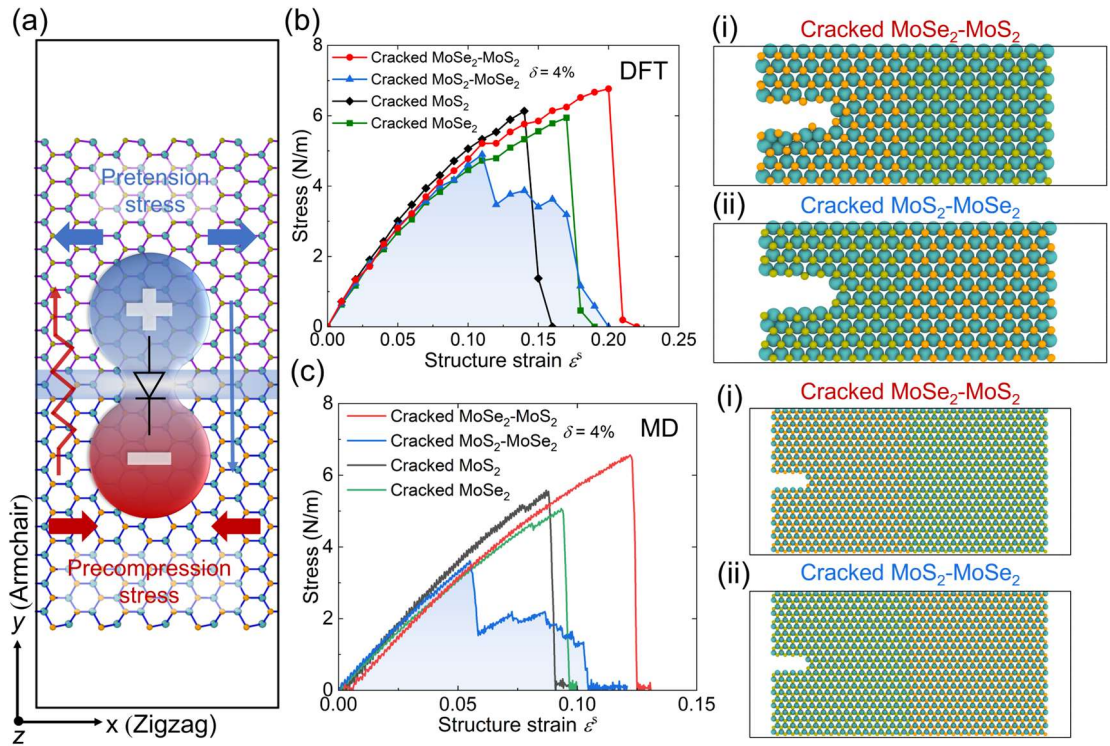

FIG. 1. Fracture nonreciprocity in TMD heterostructures. (a) Schematic of directional fracture behavior

in TMDs lateral heterostructures (made by vesta[20]), induced by interfacial lattice mismatch analogous to a diode-like response. (b, c) Comparison of stress-strain curves for four different systems with pre-existing cracks calculated via (b) quasi-static DFT and (c) MD simulations based on MACE-FT. Here, (b and c)-(i) cracked $MoSe_2$-$MoS_2$ and (b and c)-(ii) cracked $MoS_2$-$MoSe_2$, whereas $MoS_2$ and $MoSe_2$ represent pure monolayers with the same crack geometry (see SM Fig. S3).

*Molecular dynamics driven by a dedicated MLP.*—Ab initio molecular dynamics (AIMD) can, in principle, capture finite-temperature fracture, but its cost restricts accessible length and time scales. To validate fracture nonreciprocity dynamics at finite temperature, we developed a machine-learning potential (MLP) within the MACE framework[21] for the $MoS_2$-$MoSe_2$ lateral heterostructure as a representative system. The MLP was trained on DFT energies and forces to reproduce the underlying potential energy surface using equivariant message-passing graph neural networks, enabling large-scale MD at near-*ab initio* accuracy and therefore allowing us to probe crack-tip dynamics at finite temperature beyond the reach of direct AIMD.

We fine-tune the pretrained foundation model MACE-MPA-0[22] on a reference dataset comprising 993 DFT-labeled total energies and atomic forces for structures sampled from extensive MD simulations generated with MACE-MPA-0. The dataset encompasses $MoS_2$, $MoSe_2$ and their heterostructures with pre-existing cracks, as well as pristine periodic monolayers, all sampled under uniaxial tension along the armchair direction (see SM for sampling protocols). Fig. S2 compares total energies, atomic forces, and stresses across the training, validation, and test sets predicted by the fine-tuned model (MACE-FT) and pre-trained MACE-MPA-0 (MACE-MP) with DFT benchmarks. The corresponding root-mean-square errors (RMSEs) of energy, forces, and stress are below 5 meV/atom, 102 meV/Å, and 0.2 N/m, respectively, which are substantially lower than those of MACE-MP (energy, force, stress RMSEs of 38 meV/atom, 271 meV/ Å, and 0.5 N/m), demonstrating the effectiveness of our fine-tunning strategy and the reliability of MACE-FT for modeling fracture dynamics in TMD heterostructures at finite temperature.

Uniaxial-tension MD simulations based on MACE-FT were performed in LAMMPS[23] with 1 fs time steps, periodic boundary conditions in lateral directions, and free boundaries in the out-of-plane direction. We simulated rectangular systems containing 2349 atoms for five configurations: cracked $MoS_2$, cracked $MoSe_2$, cracked $MoSe_2$-$MoS_2$, cracked $MoS_2$-$MoSe_2$, and center-cracked $MoSe_2$-$MoS_2$ (SM Fig.S4). All systems were equilibrated at 300 K and 0 GPa in the isothermal-isobaric (NPT) ensemble for 100 ps to relax lateral residual stress. Afterwards, engineering strain was applied along the armchair direction at a fixed strain rate of 0.3 ns$^{-1}$ for 1 ns at 300K. The Nosé-Hoover thermostat[24] and barostat were respectively used to control temperature and pressure. During uniaxial loading, the lateral stress along the zigzag direction was maintained at zero. To capture stochastic fracture dynamics, five independent simulations were run for each system.

*Fracture nonreciprocity in TMD heterostructures.* — As a representative case, the $MoS_2$-$MoSe_2$ heterostructure with $\delta = 4\%$ exhibits pronounced fracture nonreciprocity in quasi-static DFT tensile simulations [Fig. 1(b)]. The mechanical response shows both toughening and weakening effects depending on crack orientation. Specifically, when the crack propagates from the $MoS_2$ side (smaller lattice constant $a$) toward the $MoSe_2$ region (larger-$a$), tensile prestrain at the interface reduces the resistance to bond rupture, resulting in a weakening effect. In contrast, when the loading direction is reversed (i.e., from the $MoSe_2$ to $MoS_2$ side), the interfacial compressive prestrain delays bond breaking and induces a toughening effect. Such direction-dependent behavior resembles diode rectification, where an intrinsic interfacial barrier determines the preferred direction of response[25]. Room temperature MD with the machine-learned MACE-FT potential reproduces these trends with excellent agreement [Fig. 1(c)], confirming that fracture nonreciprocity persists at finite temperature. As expected, finite-temperature MD and 0 K DFT yield minor quantitative differences in the fracture stress and strain due to intrinsic dynamic effects [Fig. S11], where thermal fluctuations advance crack nucleation and propagation relative to the quasi-static limit.

Guided by this robustness, we extend the DFT analysis across all 15 TMD heterostructures to establish the generality of fracture nonreciprocity. For all lattice-matched heterostructures ($\delta = 0\%$), uniaxial stress-strain curves remain nearly identical regardless of crack location [SM Fig. S9(a)], indicating reciprocal fracture behaviors. In contrast, all systems with finite $\delta$ exhibit pronounced directional dependence [SM Fig. S9(b-d)], demonstrating that lattice mismatch is both necessary and sufficient to induce fracture nonreciprocity in two-dimensional heterostructures.

*Modulation of fracture nonreciprocity.* — Having established lattice mismatch universally induces fracture nonreciprocity in TMD heterostructures, we now quantify how $\delta$ systematically controls its magnitude. The magnitude of fracture nonreciprocity is governed primarily by lattice mismatch through its modulation of interfacial prestrain states, which directly determine crack-tip bond-breaking thresholds. Quantitative analysis shows that the bidirectional toughness modification intensifies with increasing $\delta$ [Fig. 2(a)]. Here, we define the toughness variation rate as $K_v = (K_{HS}-K_{TMD})/K_{TMD}$, where $K_{HS}$ is the toughness of the heterostructure for a specified crack orientation and $K_{TMD}$ is that of corresponding pristine monolayer. The strongest effect occurs at $\delta = 10\%$, where the cracked $WTe_2$-$WS_2$ system exhibits a ~98% toughness enhancement relative to cracked $WTe_2$ [Fig.S10], exceeding the ~50% gains reported for topological defect engineering in graphene [26]. These results indicate that introducing the pre-crack on the side with larger lattice constant effectively leverages mismatch-induced compressive prestrain to shield the smaller-$a$ material, substantially delaying fracture initiation. This strategy increases $K$ through intrinsic interface strain fields without requiring precise nanoscale control over defect types and distributions demanded by defect-based approaches [5, 27].

To quantify this directional asymmetry, we introduce the toughness mismatch ratio $K_m = (K_{max} - K_{min}) / K_{max}$, in conceptual analogy with the rectification ratio in a photodiode[28], where $K_{max}$ and $K_{min}$ are the larger and smaller fracture toughness measured for opposite crack propagation directions, respectively. The dependence of $K_m$ on lattice mismatch $\delta$ is captured by the empirical relation $K_m = 0.4 - 0.33 \times 10^{-12\delta}$ [Fig. 2(b)], with $K_m$ reaching an upper bound of 49% at $\delta = 10\%$. This quantitative scaling law establishes a direct analogy between mechanical and electronic rectification, demonstrating a tunable "fracture diode" effect governed by interfacial lattice mismatch.

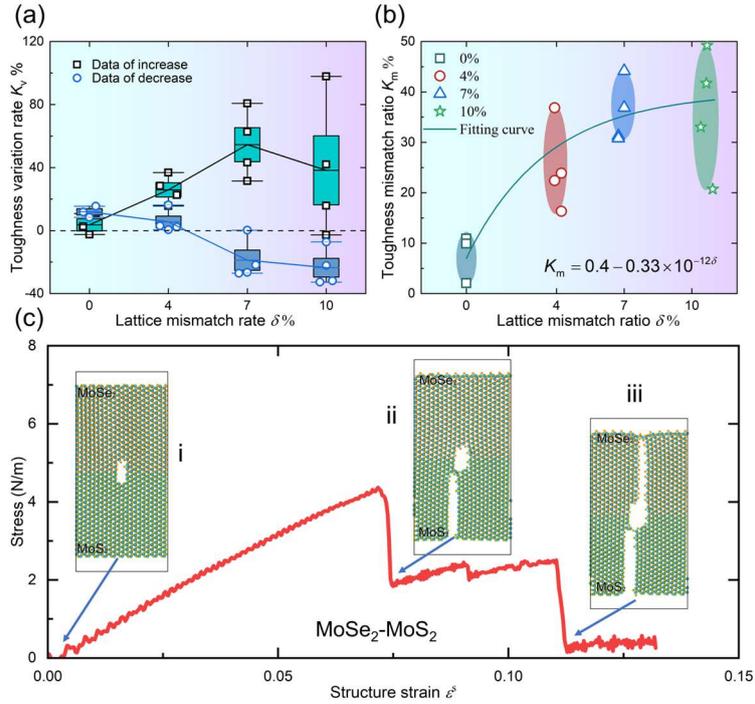

FIG. 2. The modulation of fracture nonreciprocity by lattice mismatch. (a) Systematic variation of fracture toughness ($K_v$) as a function of lattice mismatch rate ($\delta$). (b) Correlation between toughness mismatch rate ($K_m$) and $\delta$. (c) Atomistic mechanism of crack propagation in a center-cracked $MoSe_2$-$MoS_2$.

The mechanism becomes most transparent when the crack initiates at the hetero-interface. In MD simulations of center-cracked $MoS_2$-$MoSe_2$, the $MoS_2$ side (smaller $a$) fractures first (i to ii), followed by sequential failure of the $MoSe_2$ side (ii to iii) [Fig. 2(c) and Movie 1]. While the overall toughness remains comparable to the case with the crack seeded on $MoS_2$ side (see (i) in Fig.1(c)), interfacial crack initiation offers deterministic control over the propagation path: lattice mismatch-induced fracture nonreciprocity guides the crack and enforces sequential fracture across the two sides. This guided fracture sequence delays catastrophic rupture and enhances damage tolerance, enabling "fail-safe" architectures with predictable failure modes. Such directional control represents a critical design paradigm for ensuring mechanical reliability in next-generation nanoscale devices. Moreover, experimental TMD heterostructures frequently feature triangular interfaces due to their hexagonal crystal symmetry [3]. MD simulations confirm that pronounced fracture nonreciprocity persists in these configurations [SM Fig. S11 and S12], underscoring the robustness of this mechanism across realistic interface morphologies.

*Intrinsic bond failure in pristine TMDs.* — To uncover the microscopic origin of fracture nonreciprocity, we analyze chemical bond rupture processes at the atomic scale, where fracture is fundamentally governed by charge redistribution and localization of shared electron density into constituent fragments [29-31]. In monolayer TMDs with highly localized bonds [32], this yields predictable electronic redistribution patterns during rupture, thus enabling unified scaling relationships for bond failure. Furthermore, lattice-mismatched TMD heterointerfaces supply controllable strain fields at crack tips, allowing us to quantify how strain modulates these well-localized bonds and, consequently, fracture behavior.

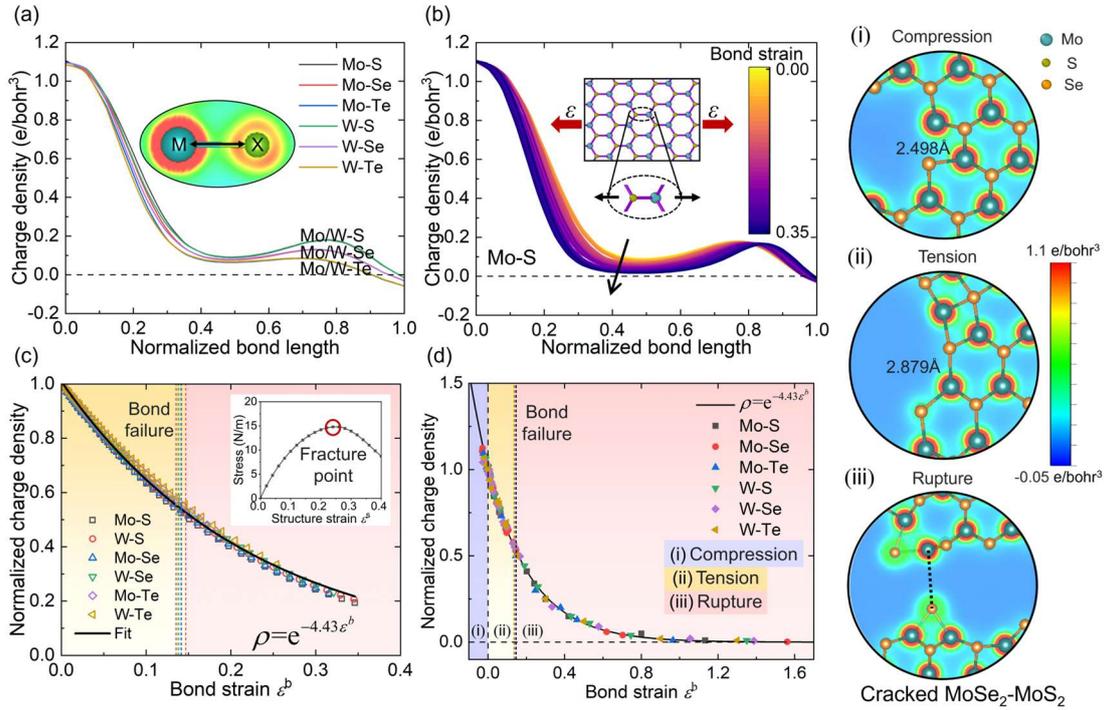

FIG. 3. Charge density analysis at crack tips. (a) Spatial distribution of charge density along the M-X bond. (b) Progressive evolution of charge density profiles along Mo-S bonds in pristine $MoS_2$ under tensile loading. (c) Universal scaling between normalized characteristic charge density and bond strain in pristine TMDs. (d) Validation of the scaling law in pre-cracked TMDs heterostructures, showing

unified behavior of M-X bonds under various states: (i) compression, (ii) tension and (iii) rupture, independent of local atomic environment.

Through systematic charge density analysis, we first characterized the electronic characteristics of M-X bonds in pristine TMDs. These bonds exhibit strongly polar covalent character [SM Fig. S14], with electron localization patterns analogous to those observed in metal-organic frameworks and complex oxides [33]. Charge density profiles along the M-X dissociation coordinate reveal pronounced asymmetry [Fig. 3(a)], indicating charge transfer from the metal (M) to the chalcogen (X). The degree of transfer decreases slightly with increasing the atomic number of X (denoted $Z(X)$), which we attribute to the concurrent reduction in electronegativity and increase in atomic radius [34].

Under tensile loading along the armchair direction, the charge density along an M-X bond evolves systematically. As the bond elongate, the charge density in the bonding region decays toward zero, signifying progressive localization of electrons onto the fragments as rupture approaches. This trend persists even after macroscopic stress begins to soften and failure occurs [Fig. 3(b) and SM Fig. S15]. To quantify the onset of rupture, we track the minimum charge density along the dissociation coordinate and, within the normalized bond length window of 0.4 to 0.6, identify the critical point at which the shared electron density collapses onto constituent fragments, marking bond rupture initiation [35, 36].

A pivotal trend emerges upon normalizing the charge density relative to its pristine value. We define the normalized charge density as $\rho \equiv \rho(l) / \rho(l_0)$, where $l_0$ and $l$ are the M-X bond lengths in undeformed and strained states, respectively. Under compression $\rho > 1$, while under tension $\rho < 1$. As shown in Fig. 3(c), $\rho$ follows a universal exponential scaling with the bond strain $\varepsilon_b$:

$$\rho = e^{-4.43\varepsilon_b} \qquad (1)$$

where $\varepsilon_b = (l - l_0) / l_0$ donates bond strain. This relation holds across all M-X bond types, independent of chemical composition and local environment. The critical charge density for rupture is determined from the peak stress in the uniaxial stress-strain curve of the pristine crystal, where the charge density at the failure strain defines the rupture threshold (insets of Figs. 3(c) and SM Fig. S7). Evidently, the normalized charge densities and $\varepsilon_b$ across different bonds show remarkable consistency, with systematic variations correlating with $Z(X)$: larger $Z(X)$ leads to earlier failure. This explains the observed decrease in fracture strength for both perfect and pre-cracked TMDs as $Z(X)$ increases [SM Tables. S1 and S3].

Crucially, this scaling law remains valid for M-X bonds in pre-cracked heterostructures, encompassing compressed, stretched, and failed states [Fig. 3(d)]. For instance, monitoring Mo-Se bonds at crack tips in cracked $MoSe_2$-$MoS_2$ shows a progression from compression to stretching and ultimately rupture, yet the normalized charge density follows the same exponential relation (Eq. (1)). Irrespective of the bond type or local environment (including the presence of interfaces and defects), the charge density evolution adheres to the unified scaling law. This universality indicates that this relation captures a fundamental aspect of bond rupture in TMDs and thus provides a quantitative framework for predicting direction-dependent fracture.

Within this framework, fracture nonreciprocity emerges as a mechanical analogue of electronic rectification: prestrain-induced asymmetry in charge density at crack tips sets directional bond-breaking thresholds, allowing cracks to propagate preferentially toward the side with lower charge density. Remarkably, the exponential scaling (Eq.(1)) mathematically parallels Shockley relation between carrier density and applied bias in semiconductor junctions[37], but underlying physics differs, with lattice mismatch rather than doping creates the interfacial asymmetry.

*Atomistic failure analysis at the crack tip.* — To map lattice mismatch to fracture nonreciprocity, we analyze bond configurations at crack tips in detail. Fig. 4(a) presents the evolution of M-X bond strain at crack tips during tensile loading. For heterostructures without lattice mismatch (e.g., cracked $WS_2$-$MoS_2$, $\delta = 0\%$), the bond strain evolution at crack tips remains nearly identical irrespective of crack location [Fig. 4(a) i-ii and SM Fig. S16]. In contrast, for cracked $MoS_2$-$WSe_2$ ($\delta = 4\%$), lattice mismatch generates distinct prestrain states at crack tips. When the crack initiates on the $WSe_2$ side, the W-Se bonds experience pre-compression from the interface strain and require additional tension to reach the failure threshold prescribed by the charge-strain scaling law (Eq. (1)). Conversely, when propagating from the $MoS_2$ side, the Mo-S bonds undergo pre-tension that brings them closer to the rupture criterion, facilitating earlier rupture [Fig. 4(a) iii-iv]. Similar trends appear for $\delta = 7\%$ and $10\%$ [SM Figs. S17 and S18]. The universality of the scaling law ensures that these prestrain differences map directly onto bond rupture thresholds.

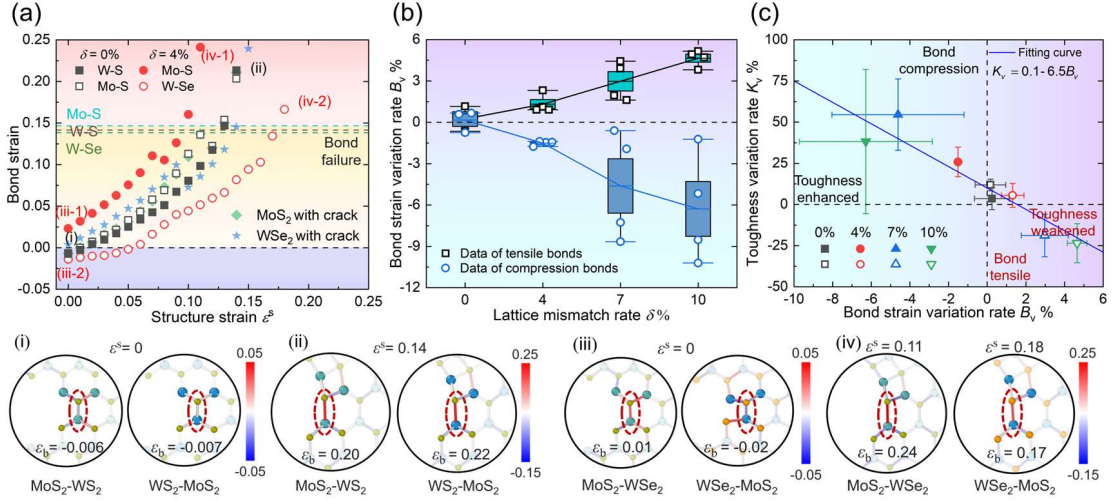

FIG. 4. Bond failure analysis at crack tip. (a) Evolution of critical bond strain at crack tip as a function of structure strain for cracked $WS_2$-$MoS_2$ ($\delta$ = 0%) and cracked $MoS_2$-$WSe_2$ ($\delta$ = 4%) with pre-cracks. Insets (i-iv) show the atomic configurations for M-X bonds at the crack tips, with bond strain magnitude represented by color mapping (made by OVITO[38]). (b) Bond strain variation rate ($B_v$) as a function of lattice mismatch rate ($\delta$). (c) Linear correlation between toughness variation rate $K_v$ and $B_v$.

The fracture evolution and propagation paths exhibit remarkable consistency across all TMD heterostructures [SM Figs. S19 and S20], indicating that fracture asymmetry originates from the prestrain states, and consequently the charge density distribution, of M-X bonds at crack tips. These asymmetric prestrain fields directly modulate bond failure thresholds, with compressive prestrain enhancing local toughness and tensile prestrain facilitating earlier rupture. Regarding this, we define the bond strain variation rate as $B_v = (B_{HS} - B_{TMD})/B_{TMD}$, where the ($B_{HS}$ and $B_{TMD}$ represent the bond stain at the crack tip in the heterostructure and the corresponding TMD counterpart, respectively. Quantitatively, $B_v$ correlates strongly with $\delta$ [Fig. 4(b)], providing a direct link between interface structure and fracture response.

More importantly, we find that $K_v$ follows an approximate linear relationship with $B_v$ [Fig. 4(c)],

$$K_v = 0.1 - 6.5 B_v \quad (2)$$

showing that the bidirectional toughening-weakening is governed by asymmetry in crack-tip bond prestrain. Together our universal scaling law for bond rupture [Eq. (2)], these results demonstrate that fracture nonreciprocity in TMD heterostructures originates fundamentally from lattice mismatch-induced asymmetric prestrain at crack tips. The consistency of this mechanism across material combinations enables direct prediction of directional fracture resistance from local strain fields, without detailed electronic structure calculations.

*Universality of nonreciprocal fracture in 2D materials.* — To establish whether interface strain engineering can generally induce fracture nonreciprocity and thus enhance nanoscale toughness beyond TMDs, we extend our study to a broader range of 2D materials. According to the 2DMatPedia database [39], there exist 6,351 2D materials (including isomers), which can be categorized into 1,432 groups based on their chemical formulas, atomic count, and space group. Excluding single-element materials leaves 602 categories capable of forming heterostructures. Structurally, 2D materials fall into four crystal families [40], hexagonal, square, rectangular, and oblique, with 78, 60, 273, and 191 groups, respectively. Restricting lattice mismatch to $0-15\%$ to ensure heterostructures stability yields possible 32,924 for hexagonal, 18,756 for square, 6,429 for rectangular, and 3,273 for oblique heterostructures [SM Fig. S21].

To assess the role of crystal symmetry systematically, we examined representative lateral heterostructures from all four families: InS-GaSe ($\delta$= 2.0%), $TaS_2$-$MoS_2$ (4.5%), and InSe-GaSe (6.0%) for the hexagonal; KBr-KCl (4.6%), TlAs-AgI (5.2%), and AgI-CuI (7.6%) for the square; SiTe-GeTe (3.1%), SiSe-SnS (7.2%) for the rectangular; and BSe-BS (6.4%) for the oblique. Consistent with our scaling law predictions in Eq. (1), all systems with $\delta > 0\%$ show clear fracture nonreciprocity, as evidenced by their uniaxial tensile responses [SM Fig. S22]. The $K_m$ values for these heterostructures increases systematically with $\delta$ across these families [Fig. 5(a)], mirroring the trend observed in TMDs heterostructures and confirming that fracture nonreciprocity transcends specific crystal symmetries.

More significantly, the linear relation between toughness variation rate $K_v$ and bond strain variation rate $B_v$ as described in Eq. (2) holds across all examined material systems [Fig. 5(b)], confirming the universality of our strain-based criterion for fracture nonreciprocity. This extensive validation establishes lattice mismatch-induced fracture nonreciprocity as a fundamental phenomenon in 2D materials. The resulting "fracture diode" effect, characterized by simultaneous strengthening in one direction and relative weakening in the reverse, offers practical opportunities for protective elements in nanoscale devices.

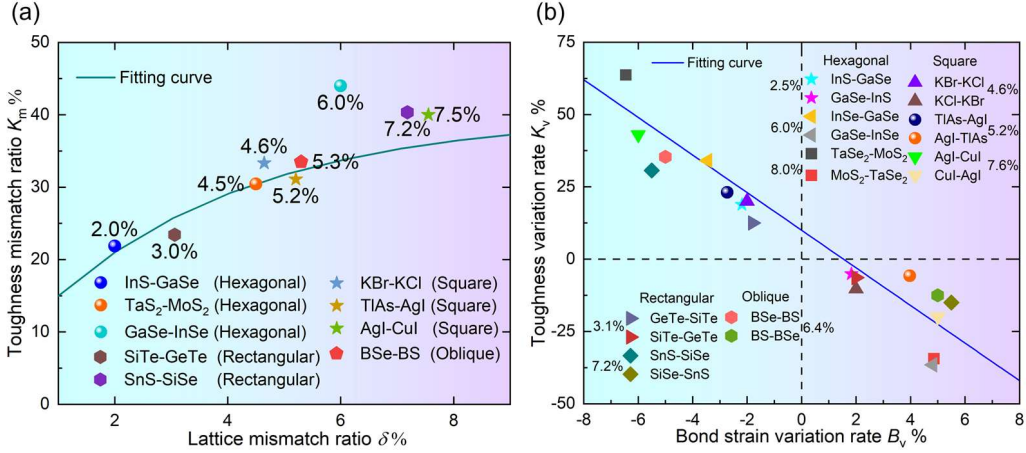

FIG 5. Universal applicability of fracture nonreciprocity across various 2D materials heterostructures. (a) Verification of fracture nonreciprocity in several representative heterostructures with $\delta$ ranging from 0 to 8%. (b) Extension of the $K_v$-$B_v$ correlation to diverse heterostructures systems, confirming the universality of the established scaling relationship (The crack is positioned in the right part of heterostructure).

*Conclusion.* — In summary, we have uncovered intrinsic fracture nonreciprocity in 2D heterostructures engineered via lattice-mismatched interfaces. Combining machine-learning MD simulations and DFT calculations, we establish a universal exponential scaling between normalized local charge density and bond strain, independent of bond chemistry or atomic environments. This law translates asymmetric crack-tip prestrain into directional fracture resistance, with toughness-mismatch ratio $K_m$ reaching 49% at lattice mismatch $\delta = 10\%$. Validation across hexagonal, square, rectangular, and oblique crystal families confirms universality.

By bridging electronic structure physics to mechanical failure, we establish interface strain engineering as a robust design principle for designing next-generation, crack-resistant nanostructures without structural degradation—enabling mechanically resilient and reliable protective elements for advanced micro- and nanoscale devices.

The data for MLP training is made available through the Zenodo [41].

*Acknowledgements*— The authors acknowledge the financial support from the National Key Research and Development Program (2024YFA1209801), the National Natural Science Foundation of China (12302140, 12325204), the China Postdoctoral Science Foundation (2023M732794, 2025T180517), the Fundamental Research Funds for the Central Universities of China (sxzy012023213), the Scientific Research Program of